\documentclass[preprint,showpacs,preprintnumbers,amsmath,amssymb, superscriptaddress]{revtex4}

\usepackage{textcomp}
\usepackage{makeidx}
\usepackage{amsmath}
\usepackage{subfigure}
\usepackage{amssymb}
\usepackage{hyperref}
\usepackage{cleveref}
\usepackage{graphicx}
\usepackage[utf8]{inputenc}
\usepackage{float}
\usepackage{color}
\usepackage{epsfig}
\usepackage{eucal}
\usepackage{amsmath}
\usepackage{tabularx}
\usepackage{tabulary}
\usepackage{graphics}
\usepackage{blindtext}
\usepackage{dblfloatfix}
\begin{document}

\title{Two-dimensional modeling of a free expansion Plasma Focus, applied to the Sumaj Lauray 720-J device with and without external electrodes}
\author{Esthefano Morales Campa\~na}
\email{esthefano.morales.campana@uantof.cl}
\affiliation{Departamento de F\'isica, Facultad de ciencias básicas, Universidad de Antofagasta, Casilla 170, Antofagasta, Chile.}
\affiliation{Departamento de F\'isica, Universidad Cat\'olica del Norte, Av. Angamos 0610, Antofagasta, Chile.}

\author{H\'ector Silva Z\'u\~niga}
\email{hector.silva@uantof.cl}
\affiliation{Departamento de F\'isica, Facultad de ciencias básicas, Universidad de Antofagasta, Casilla 170, Antofagasta, Chile.}


\begin{abstract}
The dynamics of the current sheet of a Plasma Focus device is simulated with a two-dimensional model, in the radial expansion and the axial acceleration phase. The simulation considers the free expansion of the current sheet in hydrogen gas, without cathodes, and the comparison of experimental data with cathodes and without cathodes, using the parameters of Sumaj Lauray Plasma Focus 720 J.  
\end{abstract}

\maketitle

\section{Introduction}

The plasma focus (PF) is a type of plasma generation system originally developed as a fusion power device from the early 1960s. The original concept was developed by NV Filippov \cite{filippov}, who noticed the effect while working in the first pinch machines in the USSR and by Mather \cite{mather} in the US. The basic design derives from the z-pinch concept, both the PF and the pinch use large electrical currents that pass through a gas to make it ionize in a plasma and then pinch themselves to increase the density and temperature of the plasma. Most devices use two concentric cylinders and form the pinch at the end of the central cylinder and as PF can also be found only with the internal cylinder as in the work of J. González \cite{Gonzalez}. The PF device is a known source of high temperature transient dense plasmas of n $\sim$ $10^{19}$ cm$^{-3}$ and T $\sim 1$keV which in 1980 conducted an extensive study of other phenomena generated by the PF, such as ion emissions and electrons, the emission of hard and soft x-rays, using gases such as hydrogen, argon, nitrogen, among other gases, making it useful for several applications in many different fields, such as lithography, radiography, images, activation analysis, radioisotope production , etc. Being a source of dense hot plasma, intense energy rays, etc., the PF device also shows tremendous potential for applications in plasma nanoscience and plasma nanotechnology \cite{Rawat} \cite{soto}. A PF device is a type of pinch discharge, where a high voltage pulse is applied between the coaxial cylindrical electrodes at low pressure. This generates a current sheet (CS) accelerated by Lorentz's force, beginning to rise between the electrodes. When the plasma reaches the upper end of the internal electrode (anode), the movement continues towards the center focusing the plasma in a small region forming a column of high density and temperature (pinch phase). At this point the plasma column collapses releasing beams of ions, electrons, hard and soft x-rays, or neutrons when Deuterium or Tritium is used as the discharge gas. These devices use a switch called spark-gap, which consists of two conductive electrodes separated by a space normally filled with gas that allows the circuit to close by means of a current arc activated by a high voltage pulse generated by an auxiliary circuit called trigger which discharges the energy from the capacitor bank to the vacuum chamber \cite{Silva}. The characterization of the discharge can be done with photo scintillator multicolors, which through the photoelectric effect and a process of multiplication by secondary emission is able to capture the x-rays and neutrons emitted by the plasma, another type of detector is the collector probes of charge like the faraday cups, which are capable of detecting the charged particles generated in the plasma and a mass spectrometer can also be used. The mass spectrometer is a device that is used to separate by means of electric and / or magnetic fields, ions that have a different charge-to-mass ratio within a sample in order to identify them. The description of the movement of the plasma comes from the decade of the 70s \cite{Potter} \cite{Maxon}, one of the most used models to describe the dynamics and compression of the plasma in a discharge of the PF device is the snow-plow one-dimensional model of Sing Lee \cite{Lee1} \cite{Lee2}. This type of model does not describe the temporal part of each process as the breakdown, lift-off and axial phase. However, two-dimensional models can be found that do not imply greater additional complexity, which incorporate the length of the insulator and allow obtaining information more adjusted to the experimental reality in small plasma focus devices. To study the times involved and the dynamics of CS in the early phases of the PF discharge, we have developed a two-dimensional model similar to that developed by González. We have implemented it in FORTRAN code, to simulate the operating conditions of the SUMAJ LAURAY PF device (E = 720 J), available at the University of Antofagasta.

\section{Model description and experimental set-up}\label{section2}
This model considers a plasma cylinder with cylindrical geometry that expands freely, without an external electrode (cathodes). We assume that the CS has an infinitely thin thickness and infinite conductivity, the cylinder is composed of the lateral cylindrical surface and an upper ring disc.

In the phase of radial expansion and radial-axial acceleration, the expansion of the CS is proposed, which is attached to the insulator (after the breakdown phase) and is driven by the Lorentz force $(\textbf{J}x\textbf{B})$ causing the radial expansion of the cylindrical surface and axial movement outward of the annular disc. In the radial direction, it sweeps the gas with an efficiency $fm_{r}$ and in the axial direction with an efficiency $fm_{z}$. The terms $fm_{r}$ and $fm_{z}$ are the mass fraction swept in the radial and axial direction due to the displacement of the cylindrical surface and the annular disk respectively.

\begin{figure}[H]
\centering{
\includegraphics[scale=.92]{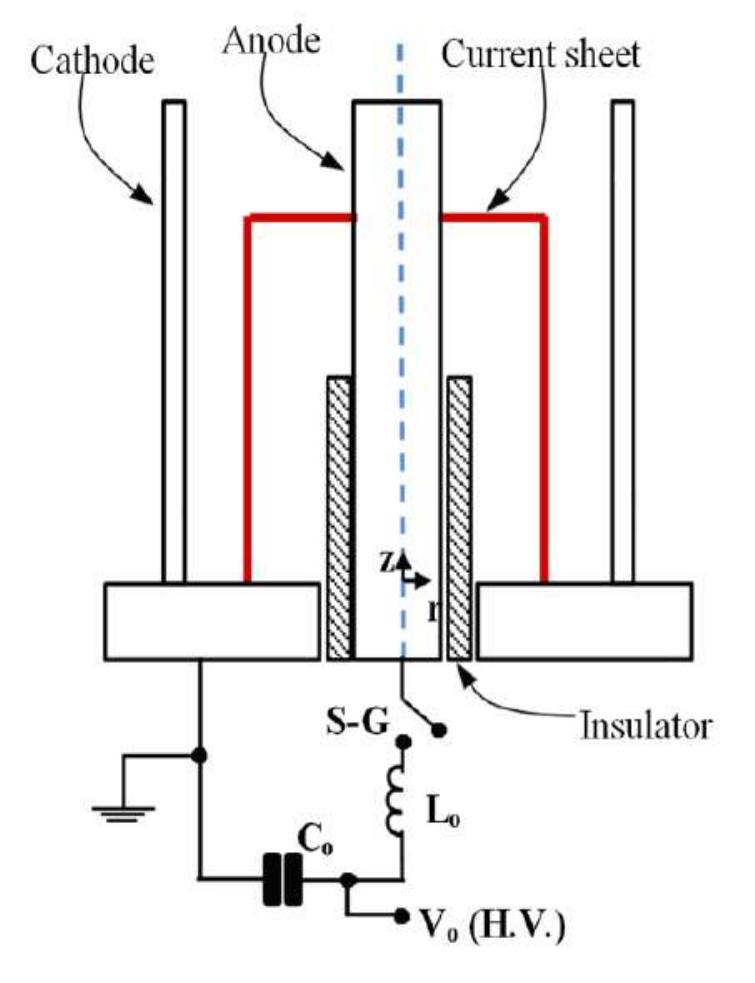}}
\caption{Two-dimensional model of plasma focus. Expansion of the plasma cylinder in a plasma focus device, Breakdown and lift-off, radial expansion and radial-axial acceleration phase, the current sheet is represented in axial phase.}
\label{fig1}
\end{figure}

\begin{figure}[H]
\centering{
\includegraphics[scale=.7]{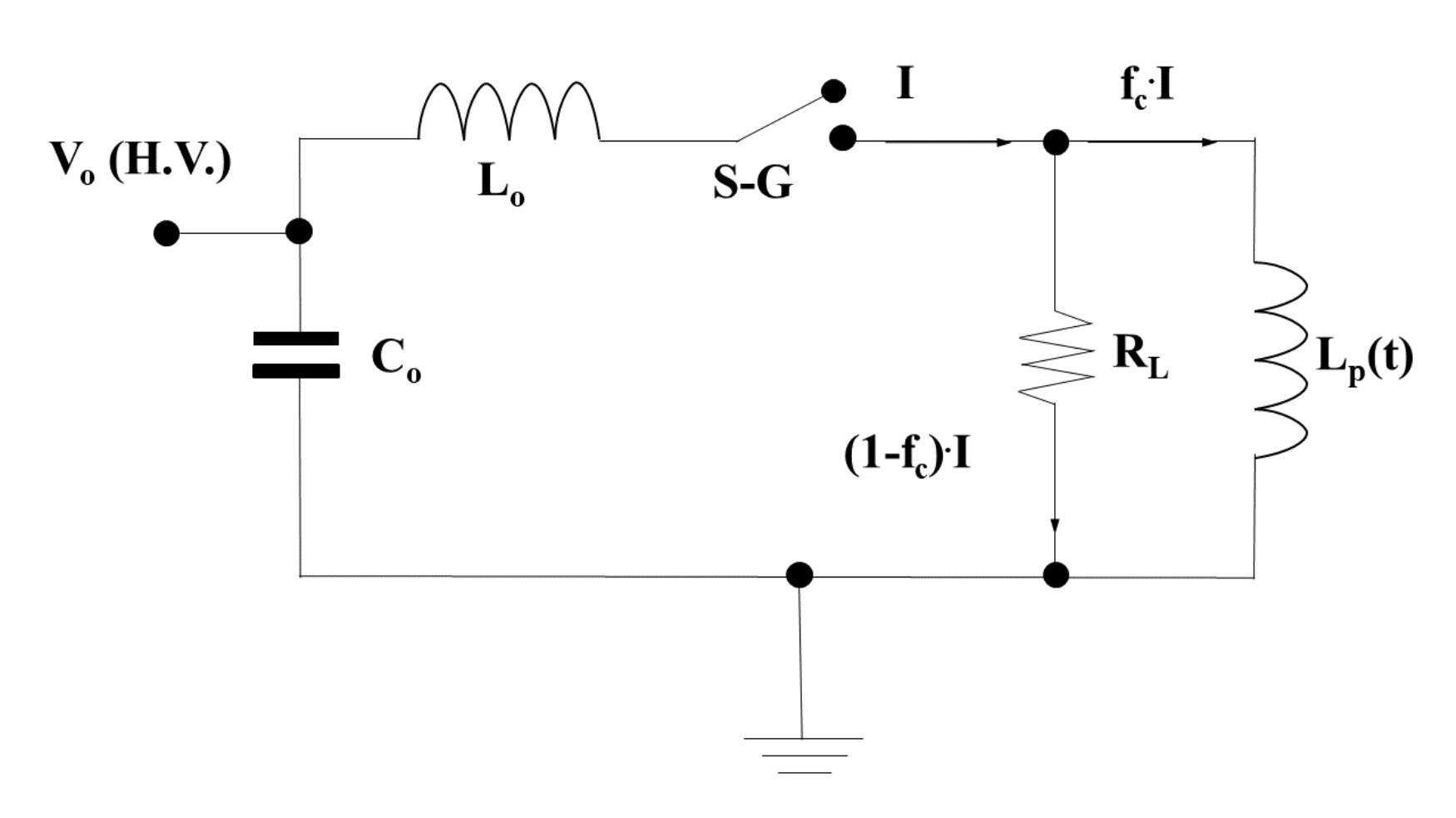}}
\caption{The figure shows the general equivalent circuit of plasma focusing device, the current factor $f_{c}$ is the current that is used to generate the current sheet, the rest of it $(1-f_{c})$ is the current leakage.}
\label{fig2}
\end{figure}

Figure \ref{fig1} shows a representative general scheme of the Mather type plasma focus with the outer and inner electrodes. The capacitor bank is represented by the capacity Co, the inductance of the system by Lo, the spark-gap by S-G. The internal electrode is the anode and the external electrodes are the cathodes, separated by the cylindrical insulator.

Figure \ref{fig2} shows the general equivalent circuit of the plasma focus device. The inductance formed by the plasma current sheet and the electrodes is represented as a time-dependent inductance Lp (t). We assume that only a fraction of the total current (I) flows through the CS ($f_{c} \cdot I$), due to the leakage of current in another region of the discharge. Then, the current factor $f_{c}$ is the term that determines the fraction of current flowing through the CS. The processes that generate the leakage current, (1-$f_{c}$)$\cdot$I, are represented by the resistance RL.

\section{Model equations}

The movement equations are obtained for the axial direction and for the radial direction of the plasma cylinder, two acceleration equations of the CS are obtained, in addition two equations are obtained for the variation of the axial ($M_{z}$) and radial mass ($M_{r}$) with respect to time. The equation describing the plasma focus circuit is taken as an LC circuit with an additional voltage that delivers the effect of the simulation of the Spark-Gap plus the electrical breakdown of the plasma. Using the magnetic vacuum permeability ($\mu_{0}$), the density of the gas ($\rho$) and the radius of the anode (a), we can write the following equations:
\begin{equation}\label{eqn:1}
\normalsize{\frac{d}{dt}[(L_{0}+f_{c} L_{p}(t))I(t)]=(V_{0}-V_{sg}(t))-\int_{0}^{t} \frac{I(t')}{C_{0}}dt'}
\end{equation}
\begin{equation}\label{eqn:2}
\normalsize{\frac{d}{dt}{\left\lbrace{M_{z}(t)\frac{dz}{dt}}\right\rbrace}=\frac{\mu_{0}}{4\pi}ln(\frac{r(t)}{a})(f_{c} I(t))^{2}}
\end{equation}
\begin{equation}\label{eqn:3}
\normalsize{\frac{d}{dt}{\left\lbrace{M_{r}(t)\frac{dr}{dt}}\right\rbrace}=\frac{\mu_{0}}{4\pi}\frac{z(t)}{r(t)}(f_{c}I(t))^{2}}
\end{equation}
\begin{equation}\label{eqn:4}
\normalsize{\frac{dM_{z}}{dt}=\pi\rho(r(t)^{2}-a^{2})f_{mz}\frac{dz}{dt}}
\end{equation}
\begin{equation}\label{eqn:5}
\normalsize{\frac{dM_{r}}{dt}=2\pi\rho r(t)z(t)f_{mr}\frac{dr}{dt}}
\end{equation}
where,
\begin{equation}\label{eqn:6}
\normalsize{L_{p}(t)=\frac{\mu_{0}}{2\pi}ln(\frac{r(t)}{a})z(t)}
\end{equation}
and, the Spark-Gap voltage is obtained of Bruzzone model \cite{Bruzzone}. The initial conditions assume a thin plasma sleeve on the insulator, that begin in rest. With the volume of this plasma sleeve and the gas density, we obtain the initial mass of $M_{r}$ and $M_{z}$. The initial current and its derivative are zero. The initial conditions of the variables for higher derivatives can be obtained from the proposed equations.


\section{Sumaj Lauray Plasma Focus}\label{section3}
The experimental data are obtained from the plasma focus device of the University of Antofagasta, called Sumaj Lauray (405-1125 [J]). With a Rogowski coil, the arrival time of the CS at the end of the anode can be measured. To measure the arrival of the CS, at different axial positions, anodes of different lengths were used. Hydrogen was used as a filling gas in the PF discharge, with a pressure range of 1 to 5 [Torr]. The statistic of ten measurements, for each anode length and gas pressure, was considered in the analysis of experimental data. To simplify the simulation boundary conditions, the PF was operated without the cathode bars and also experimental data were taken with the cathode bars. The main features and working conditions of this PF device are presented in the following table:

\begin{table}[htbp]
\begin{center}
\caption{Experiment and model data table.}
\begin{tabular}[t]{ccc}
\hline
Symbol & Parameter & Value\\
\hline 
$E$ & Plasma focus energy & $720$ J \\
\hline
$I_{0}$ & Maximum current & $109,1$ kA\\\hline 
$C_{0}$ & Capacitor bank capacity & $3.6$ $\mu F$ \\\hline
$L_{0}$ & Initial inductance & $121$ nH\\\hline 
$L_{c}$ & Base Inductance of the canyon & $12$ nH\\\hline
$a$ & Anode radius & $6$ mm \\\hline
$b$ & Cathode radio & $20$ mm \\\hline
$z_{a}$ & Insulator length & $26$ mm \\\hline 
$r_{a}$ & Insulation radius & $9$ mm \\\hline 
$p$ & Filling pressure & $1$ a $5$ Torr\\\hline 

\end{tabular}
\label{tabla:4.1}
\end{center}
\end{table}


\section{Results and discussion}\label{paf}
 
\begin{figure}[H]
\centering{
\includegraphics[scale=.51]{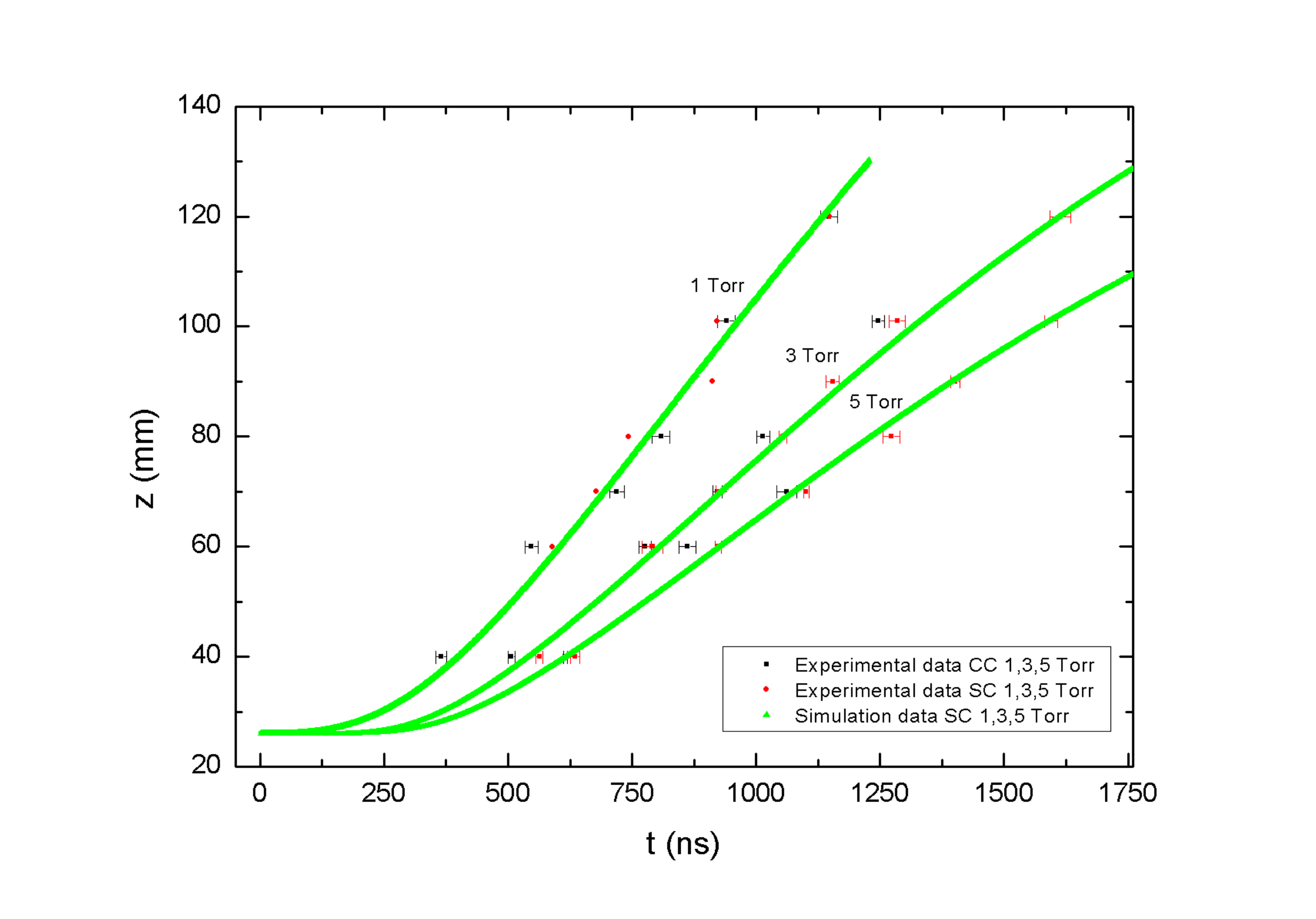}}
\caption{Experimental data with cathodes (CC) and without cathodes (SC), and simulation adjustment of the axial position, versus time, of the annular disk for pressures of 1, 3, 5 Torr.}
\label{fig3}
\end{figure}

The model presented consists of the following phases; the Spark-Gap, breakdown and lift-off phase, radial expansion and the axial phase, developing the simulation in fortran language. The results obtained are governed by the studied geometry, a straight cylindrical geometry of the plasma is established which is an idealization of the plasma movement. The initial occupied voltage is 20 kV, the voltage associated with the potential drop differs from the pressure of the filling gas, when there is more pressure in the vacuum chamber the electrical breakdown can begin before 20 kV, for the reason of that the increase in pressure makes the electrical breakdown more unstable and does not reach 20 kV. One of the experimental errors that can be treated are the measurements at a certain voltage, this instrumental (systematic) error has a variation of $\pm$ 0.5 kV, which gives us a certain margin of error in the statistics made with the 10 shots. This margin of error is identified at the position of the points in the axial position graph with respect to time. 10 shots are made for different lengths of the cannon, $dI/dt$ curves are obtained which clearly identifies the pinch, where the time of arrival at the end of the anode is obtained. The same procedure is performed for lengths of 40, 60, 70, 80, 90, 101 and 120 mm for pressures of 1–5 Torr.
The Spark-Gap voltage modeling allows the temporal adjustment of the $dI/dt$ simulation to the signal measured by the Rogowski coil in the PF device. We can estimate the current factor ($f_{c}$) from the plasma voltage $(V_{p}=d(f_{c}L_{p}(t)\cdot I(t))/dt)$, when the CS has reached the mouth of the anode. These estimates give us values of: $f_{c}=0.77$ to $0.79$ for pressures of 1 to 5 Torr. Therefore, the mass drag factors, which allowed adjusting the experimental data of Z v/s t, were: $fm_{r}$= 0.82 to 0.79, and $fm_{z}$= 0.18 to 0.21 approximately, for the same pressure range \cite{Hawat}. 
\\
Figure \ref{fig3} shows the comparison of the data of the axial position of the current sheet for pressures of 1 to 5 Torr of filling pressure, which had already been taken previously in the Plasma Focus Sumaj Lauray. The trend of experimental data with cathodes follows the curves of free expansion simulations (without cathodes), which indicates that the current sheet behaves in the same way, either a cathode model or without cathodes. This trend in the dynamics of the dynamics of CS in simulation without cathodes, compared with data with cathodes and without cathodes, clearly shows that during the process in which the plasma is generated and accelerated by Lorentz force through There is no variation of the canyon, therefore, what affects the production of the ion particle beam, hard or soft x-rays are the initial and final phases of the PF. 
 
\begin{figure}[H]
\centering{
\includegraphics[scale=.51]{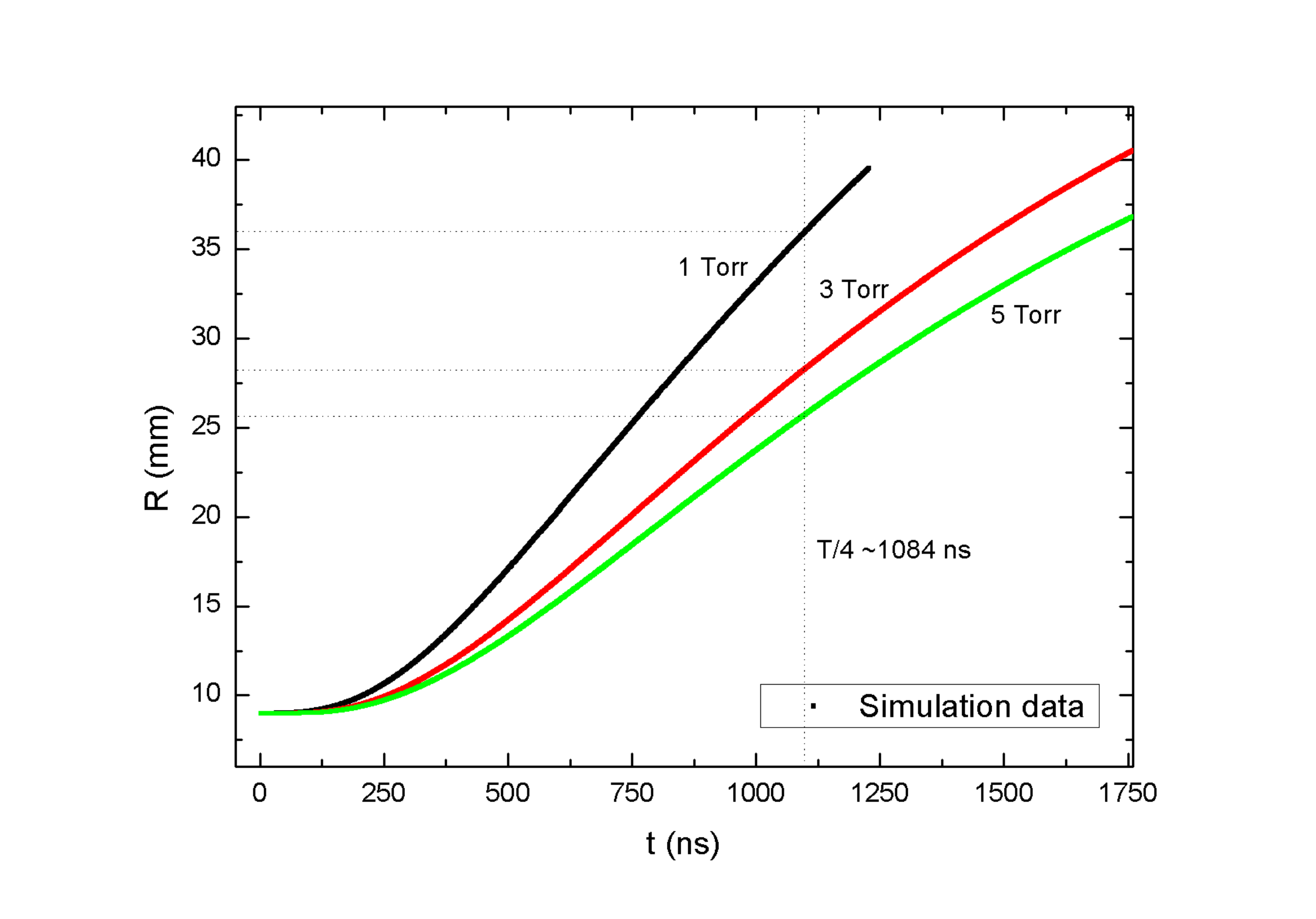}}
\caption{Radial position simulation, of the plasma cylinder, versus time. The radial range, at the current peak instant (1084 ns), is shown.}
\label{fig4}
\end{figure} 


The radial position of the CS according to the proposed model, as shown in Figure \ref{fig4}, tells us that the CS expands radially to values greater than 35 mm in radius of the plasma focus device, which indicates that for the cathode model, the plasma that passes between the cathodes could expand to these limits or simply the radial movement can be restricted by the cathodes without affecting the speed it carries. The figure shows the radii reached until the fourth period when $dI/dt=0$ for the different pressures from 1 to 5 Torr.

\begin{figure}[H]
\centering{
\includegraphics[scale=.51]{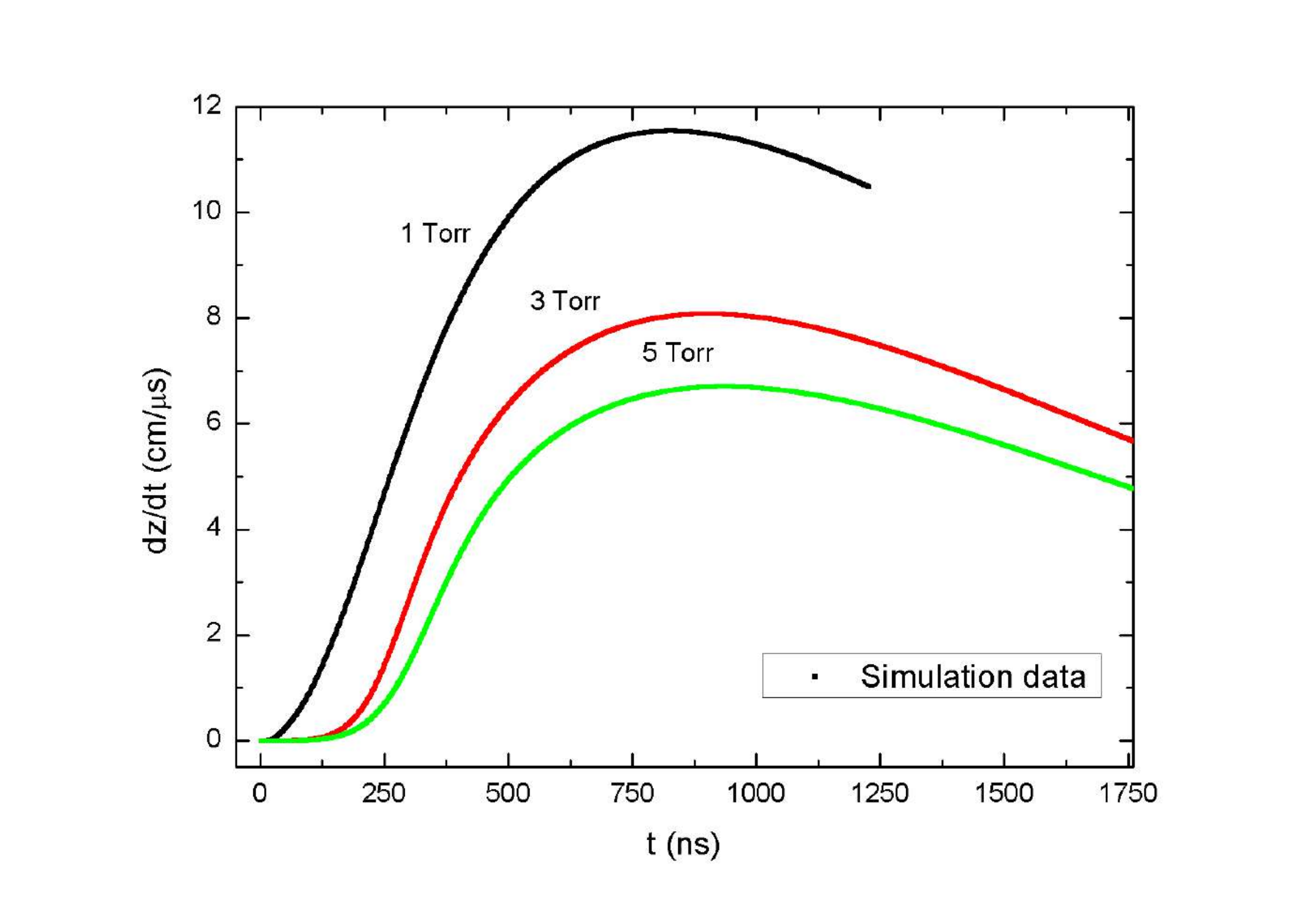}}
\caption{Simulation of the axial velocity of the CS with respect to time. For the pressure of 1 Torr, high and low energy ions are obtained with a velocity close to $10$ $cm/\mu s$.}
\label{fig5}
\end{figure}

\begin{figure}[H]
\centering{
\includegraphics[scale=.51]{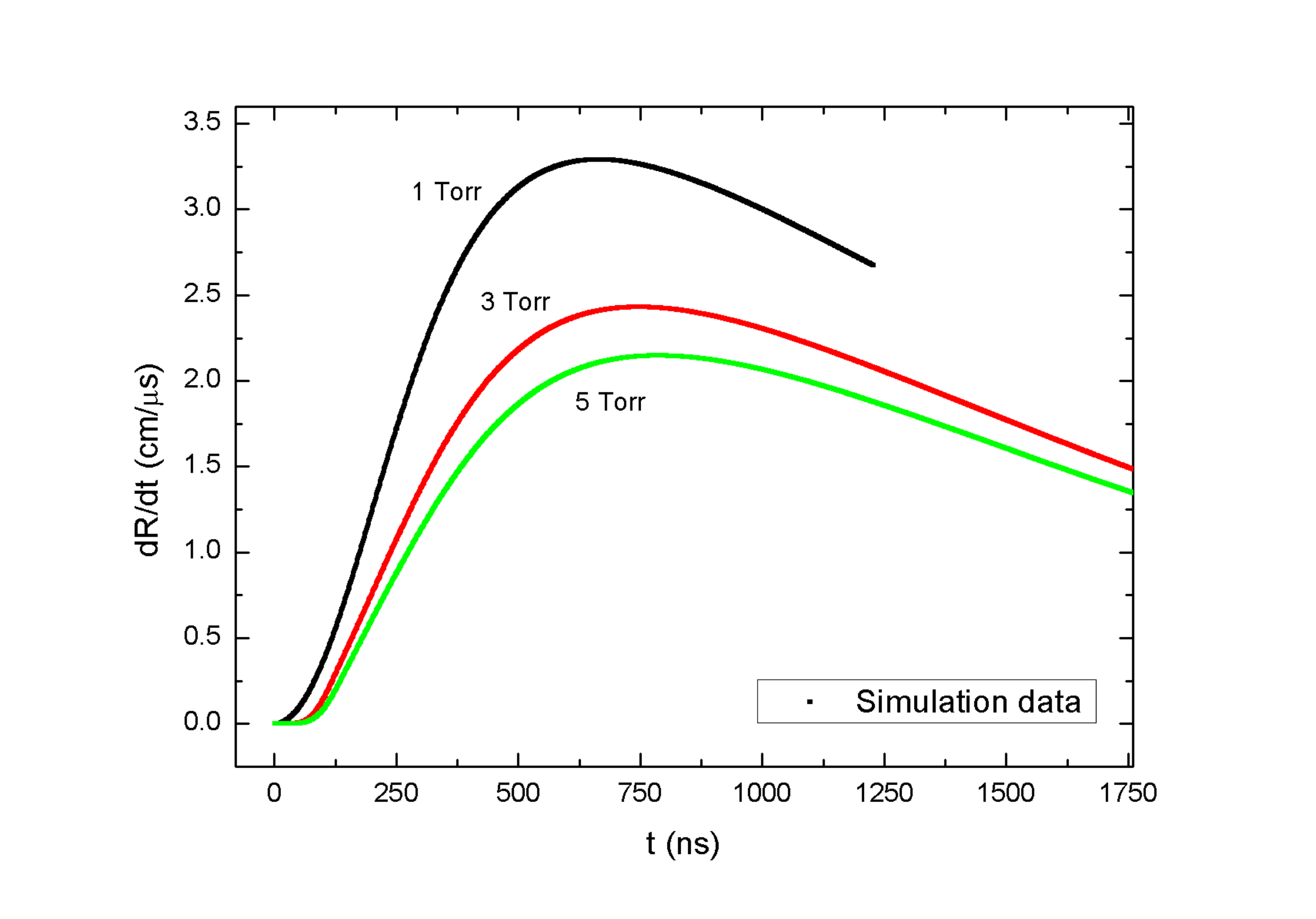}}
\caption{Simulation of the radial velocity of the CS with respect to time. The
radial velocity is one third of the axial velocity.}
\label{fig6}
\end{figure}

The speed of the CS is important for the study of the production of ions, see figure \ref{fig5}, where it is shown that for 1 Torr the speed reached close to $12$ $cm/\mu s$ at the time of $\sim 850$ $ns$, this indicates that there are less amount of mass dragged in the axial position. In Figure \ref{fig6}, it is observed that the speeds in the radial position of the system is lower than the speeds of the axial position, this is affected by the amount of mass that is dragged during the CS process.

\begin{figure}[H]
\centering{
\includegraphics[scale=.51]{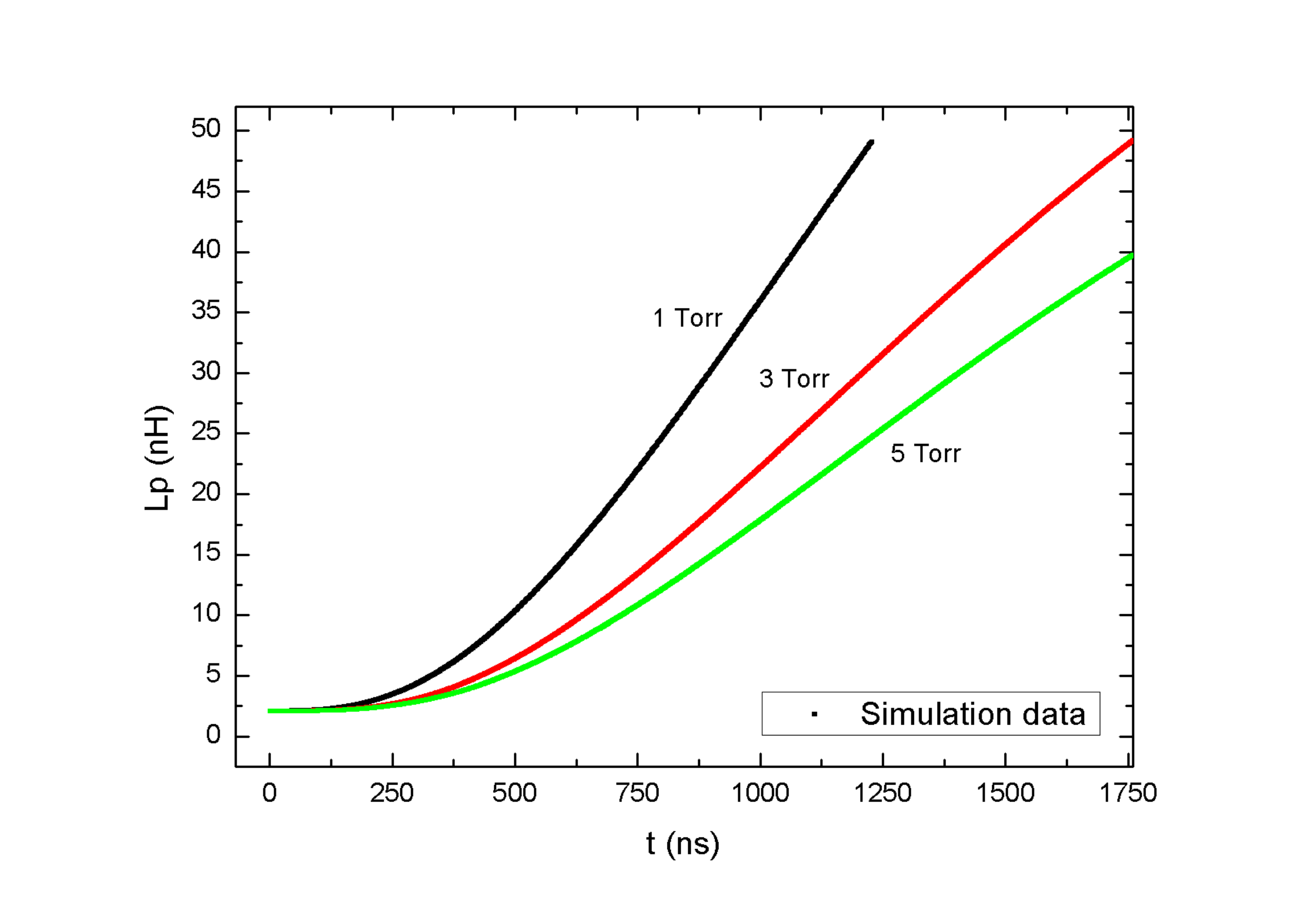}}
\caption{Simulation of the inductance of the CS with respect to time. The initial inductance is 2.1 nH independent of the filling pressure. For these curves the breakdown phase is noted where it occurs in the first hundreds of nanoseconds.}
\label{fig7}
\end{figure}

\begin{figure}[H]
\centering{
\includegraphics[scale=.51]{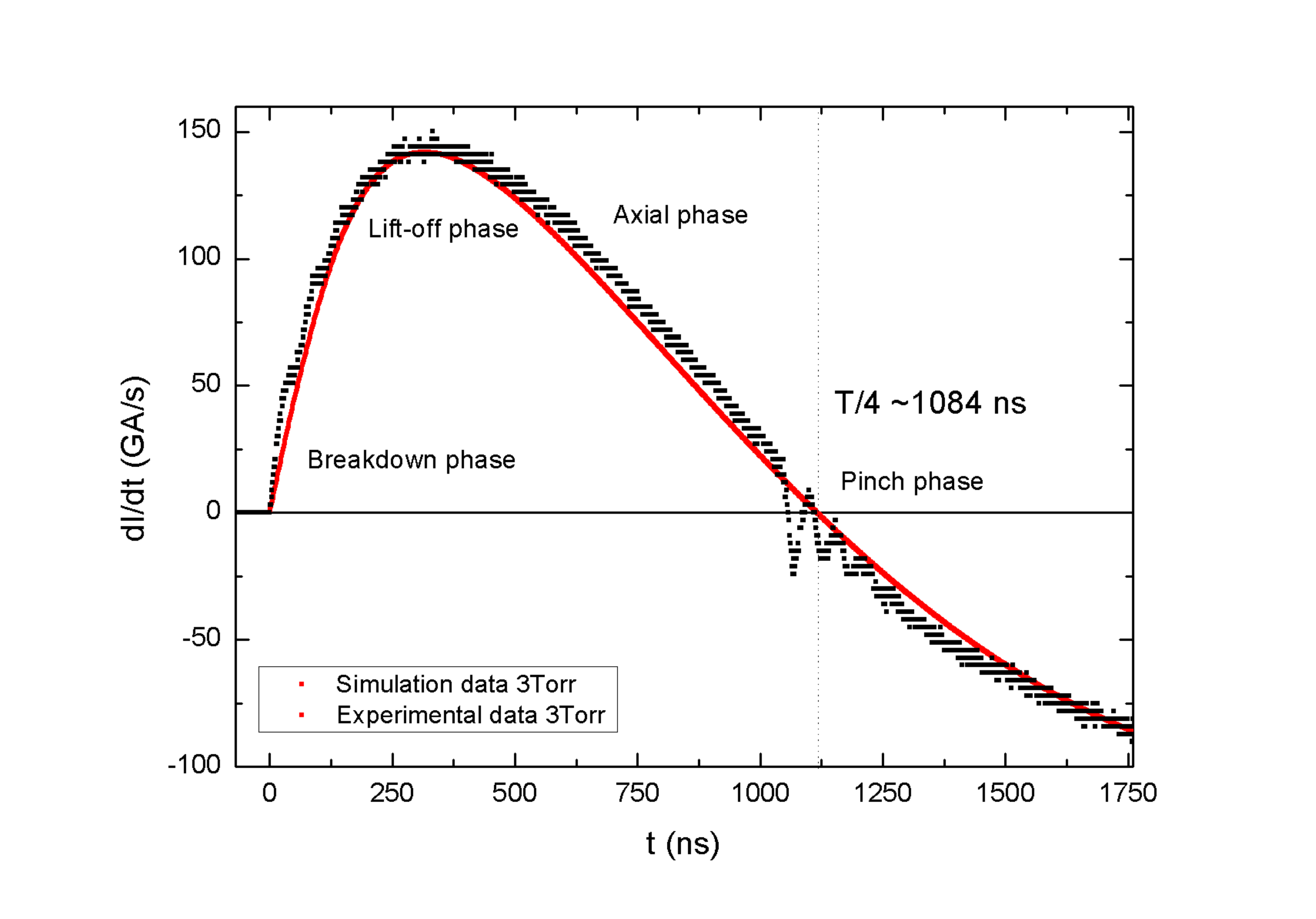}}
\caption{Simulation of the inductance of the CS with respect to time. The initial inductance is 2.1 nH independent of the filling pressure. For these curves the breakdown phase is noted where it occurs in the first hundreds of nanoseconds.}
\label{fig8}
\end{figure}

Figure \ref{fig7} shows the plasma inductance generated during the CS processes. The initial plasma inductance is 2.1 nH, this initial inductance is independent of pressure, since it only depends on the cylinder geometry of the generated plasma. The curves of the temporal derivative of the current in Figure \ref{fig8}, are observed
the simulation and experimental data for the 3 Torr pressure, which is adjusted by the current and mass drag factors.

\begin{figure}[H]
\centering{
\includegraphics[scale=.51]{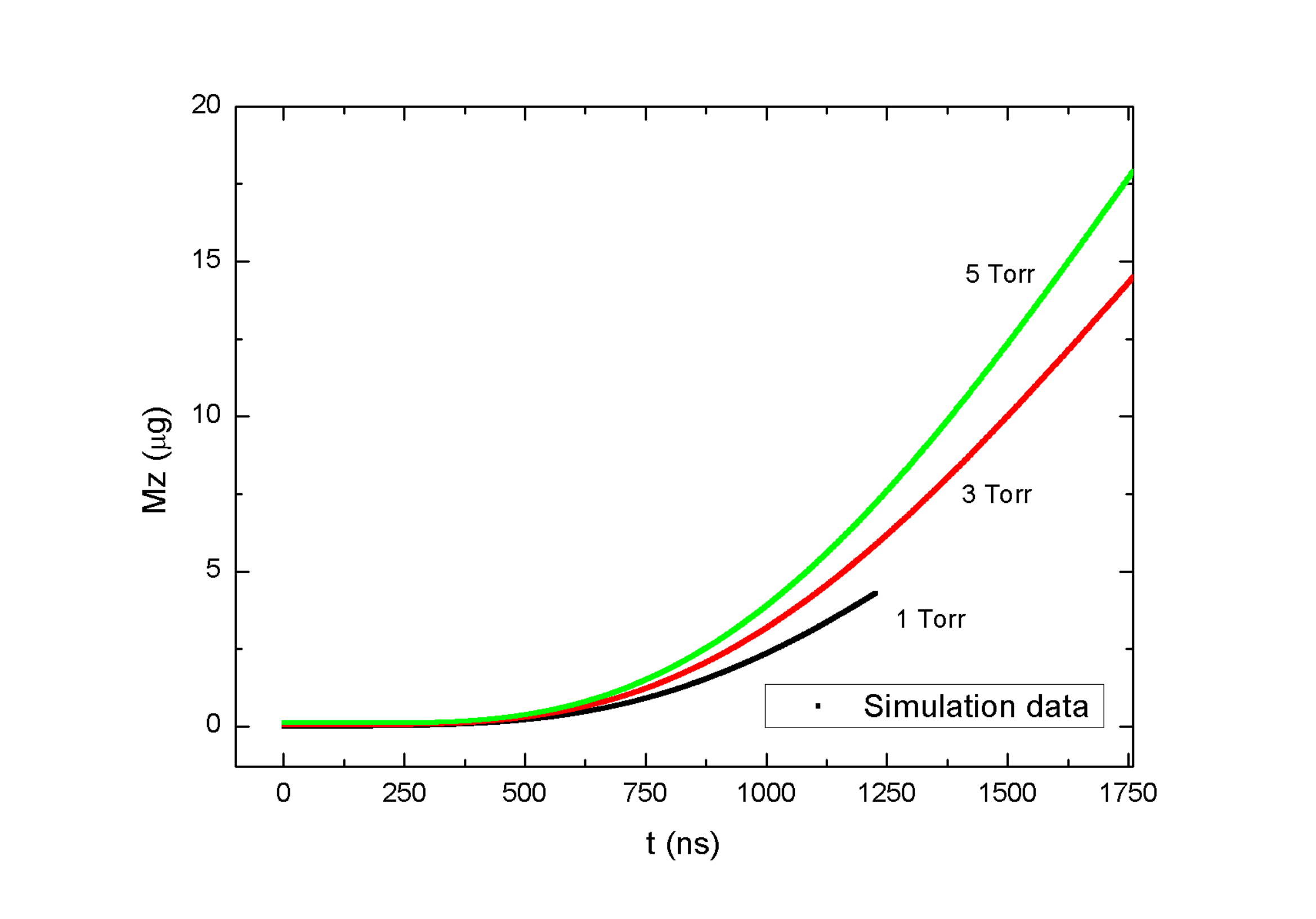}}
\caption{Simulation and experimental data of the temporal derivative of the current with respect to time. The black lines indicate the fourth period of the proposed two-dimensional model and the phases of the CS process are named.}
\label{fig9}
\end{figure}

\begin{figure}[H]
\centering{
\includegraphics[scale=.51]{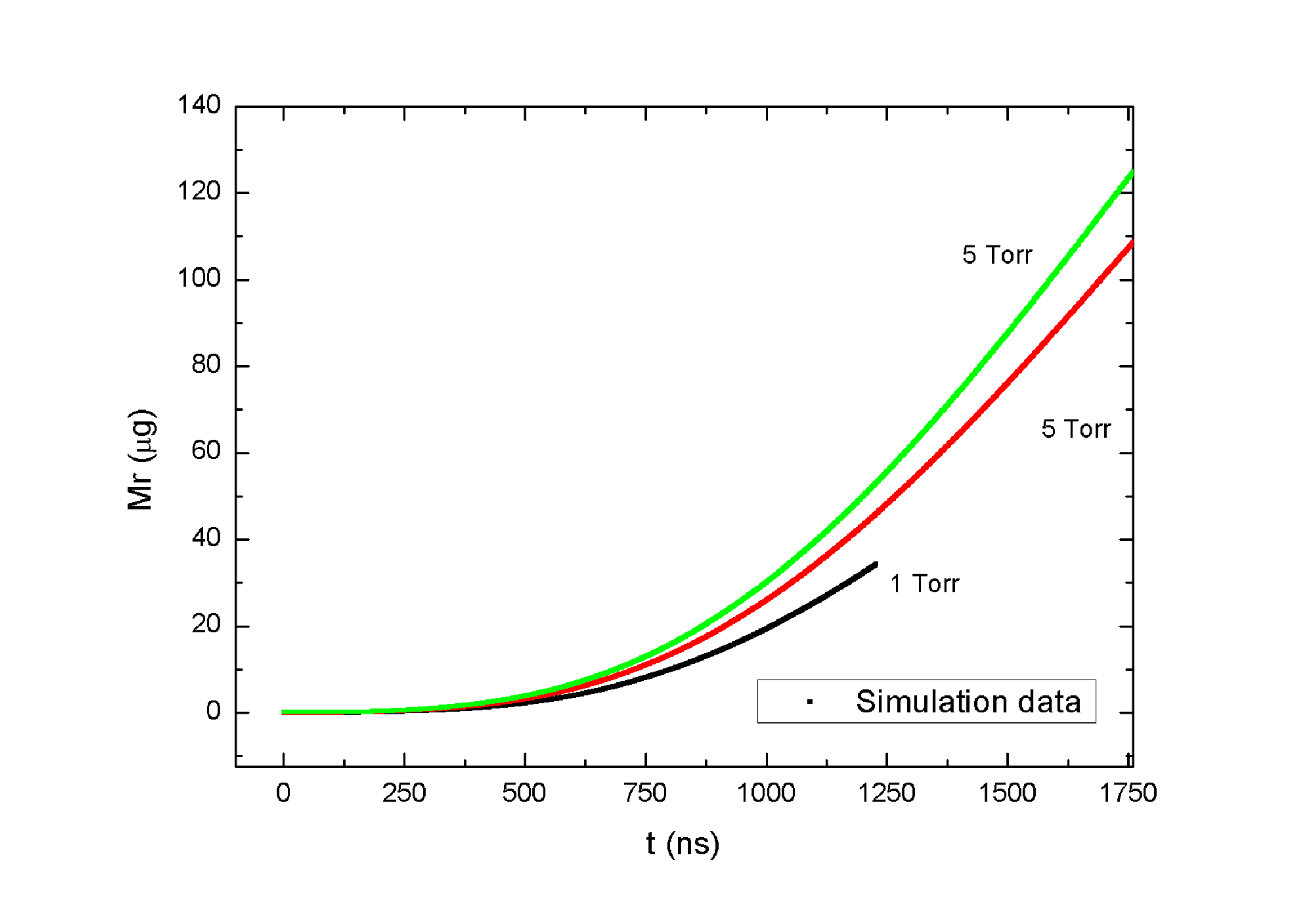}}
\caption{Simulation of the axial mass of the CS with respect to time. Mass obtained for the pressure of 1 Torr is $3 \mu g$ and for 5 Torr it is less than $24 \mu g$.}
\label{fig10}
\end{figure}

Figure \ref{fig9} shows the amount of axial mass in units of micrograms, for the pressure of 1 Torr as shown in the graph it reaches a value of 3 $\mu g$ and for the pressure of 5 Torr a value of 24 $\mu g$ is reached. In Figure \ref{fig10}, it is observed that in the radial position a greater amount of mass is obtained in the order of tens and hundreds of micrograms. This checks the values found in the radial and axial speeds in the figures shown above.

\begin{figure}[H]
\centering{
\includegraphics[scale=.51]{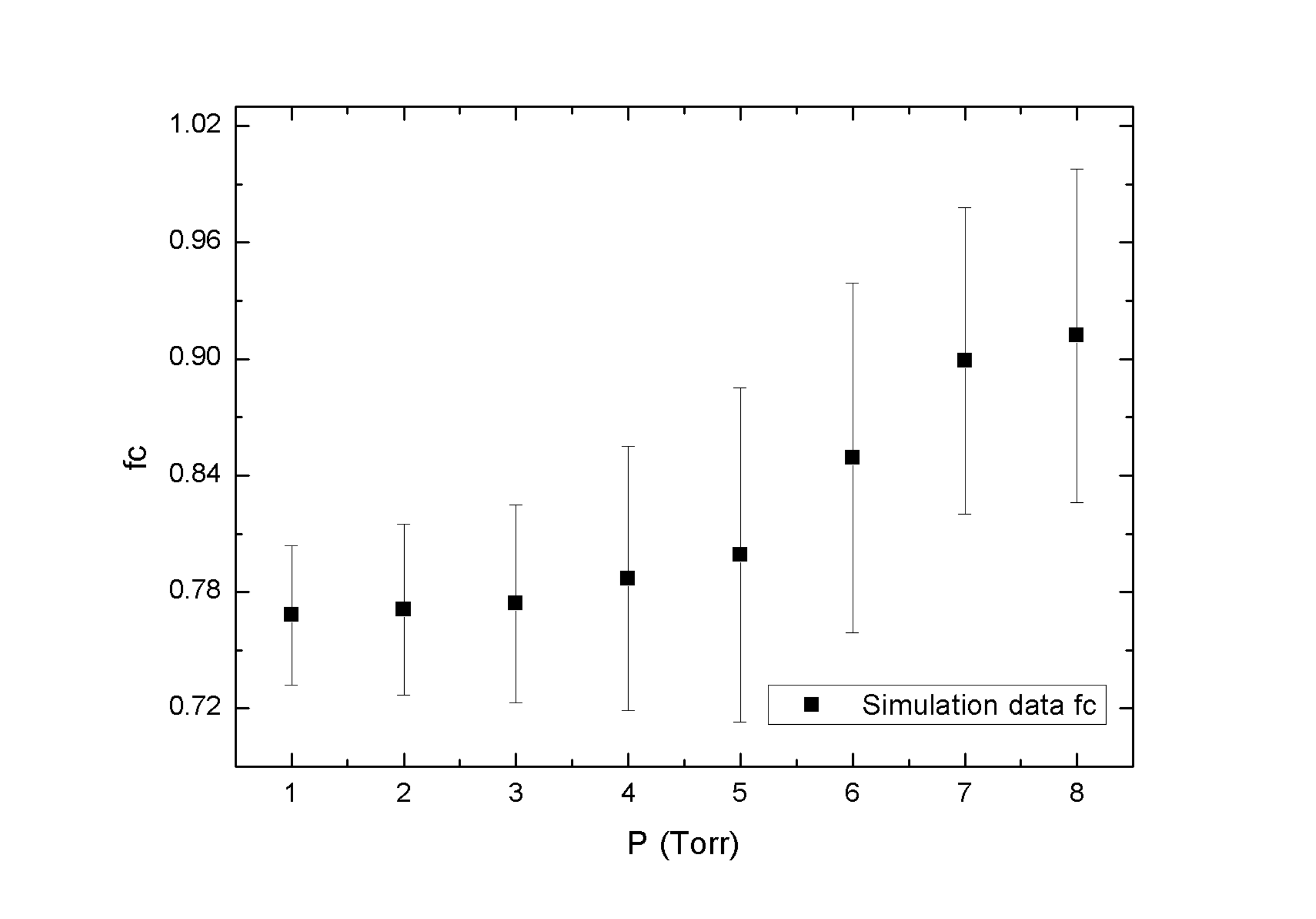}}
\caption{Current factor with respect to the filling pressure, the behavior of the points is almost linear as the pressure increases. For 1 torr $f_{c}$ is 0.768, 3 torr $f_{c}$ is 774 and 5 torr $f_{c}$ is 0.799..}
\label{fig11}
\end{figure}

\begin{figure}[H]
\centering{
\includegraphics[scale=.51]{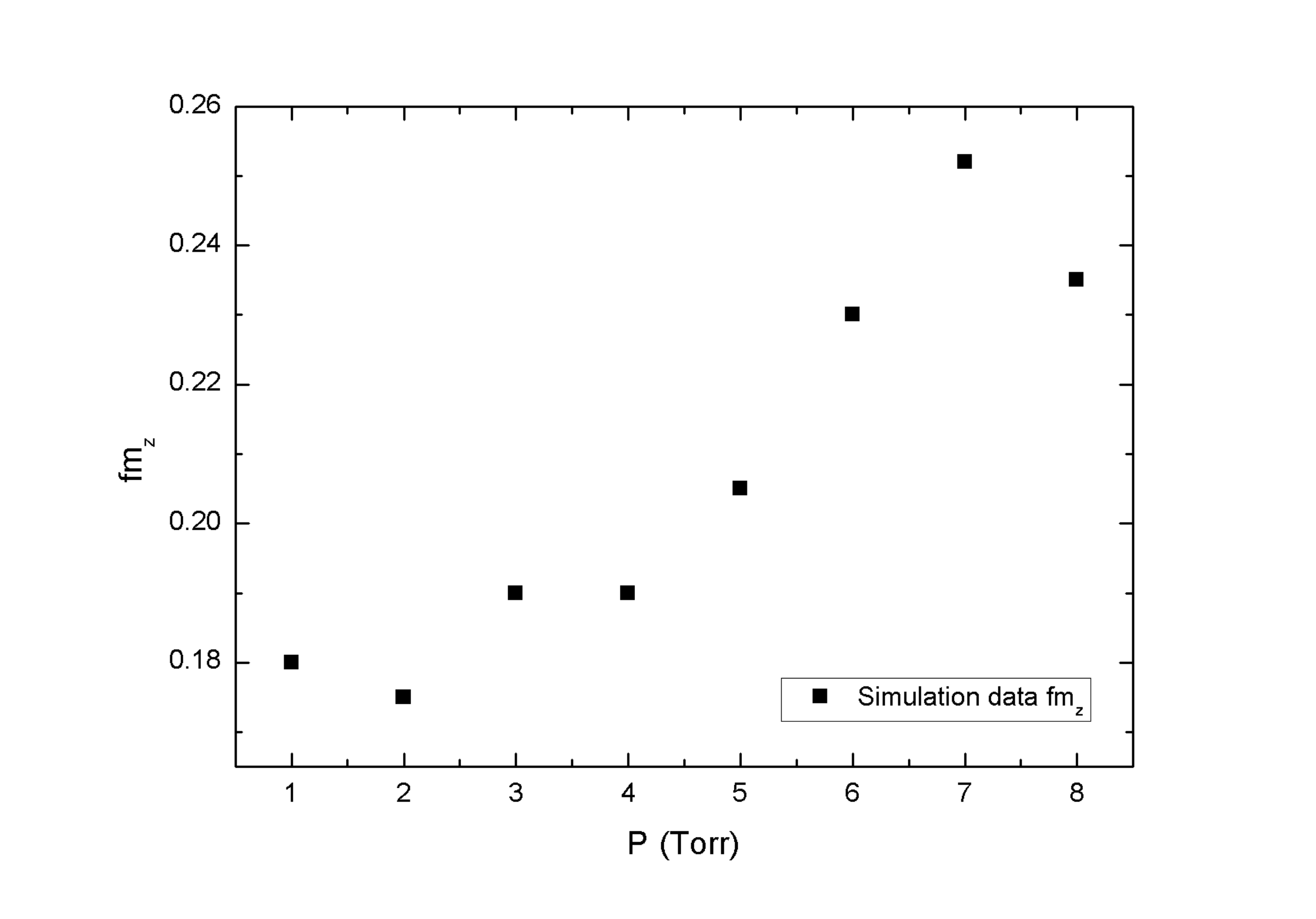}}
\caption{Axial mass drag factor with respect to the filling pressure of 1 to 8 Torr. For 1 torr $fm_{z}$ is 0.18, 3 torr $fm_{z}$ is 0.19 and 5 torr $fm_{z}$ is 0.205.}
\label{fig12}
\end{figure}

\newpage
The current factor is obtained from the experimental data semi-empirically, see figure \ref{fig11}, using the average of the 10 shots taken for each length of the cannon at each pressure. The trend of the current factor gives us the necessary information to understand that during the ionization process of the Hydrogen gas for each pressure is different, while the pressure increases the current factor tends to unity and this clearly affects the factor of axial and radial mass drag. The data used for comparison between experimental data with cathodes (CC) and without cathodes (SC), with simulation data, only the pressure range of 1 to 5 Torr is used. The axial mass drag factor is obtained after obtaining the current factor, since the current factor is obtained experimentally and among the axial and radial mass drag factors the one that infers the most in the equations is the axial drag factor. This factor is obtained from the simulation settings, compared to the least squares adjustment of the experimental data, see figure \ref{fig12}.


\section{Concluding remarks}\label{section7}
We have implemented a two-dimensional model to simulate the dynamics of the current sheet of a Plasma Focus device. This model considers only the breakdown, liff-off and axial phases of the Plasma Focus discharge, and considers the start of the discharge in the insulator. The model considers the free expansion of CS in hydrogen gas, without cathodes. The simulation was implemented with a FORTRAN code, using the parameters of the Plasma Focus device of the University of Antofagasta, Sumaj Lauray 720 J as a reference. This model provides us with a powerful tool to facilitate the interpretation of experimental data and for the design of low energy PF devices. Future work considers improving simulations and extending the model to the dense plasma phase.

\newpage

\section{Acknowledgements}
This work has been funded by FONDECYT grant 1130787
and Esthefano Morales Campa\~na thanks
the financial support of the project ANT-1856 at the Universidad de Antofagasta, Chile.

\end{document}